\documentclass{emulateapj}
\usepackage{graphicx}
\usepackage{natbib}
\usepackage{bm}
\usepackage{verbatim}

\begin{document}

\title{Vortex migration in protoplanetary disks}
\author{Sijme-Jan Paardekooper, Geoffroy Lesur and John C. B. Papaloizou}
\affil{DAMTP, Wilberforce Road, Cambridge CB3 0WA, United Kingdom}
\email{S.Paardekooper@damtp.cam.ac.uk}

\begin{abstract}
We consider the radial migration of vortices in two-dimensional isothermal gaseous disks. We find that a vortex core, orbiting at the local gas velocity, induces velocity perturbations that propagate away from the vortex as density waves. The resulting spiral wave pattern is reminiscent of an embedded planet. There are two main causes for asymmetries in these wakes: geometrical effects tend to favor the outer wave, while a radial vortensity gradient leads to an asymmetric vortex core, which favors the wave at the side that has the lowest density. In the case of asymmetric waves, which we always find except for a disk of constant pressure, there is a net exchange of angular momentum between the vortex and the surrounding disk, which leads to orbital migration of the vortex. Numerical hydrodynamical simulations show that this migration can be very rapid, on a time scale of a few thousand orbits, for vortices with a size comparable to the scale height of the disk. We discuss the possible effects of vortex migration on planet formation scenarios.   
\end{abstract}

\maketitle

\section{Introduction}
Planets are thought to form in circumstellar disks of gas and dust that are commonly found around young stars. Within these disks, planet formation requires that solid material has to grow 13 orders of magnitude in size, from sub-micron sized interstellar grains all the way towards terrestrial planets. Large-scale, long-lived vortices can play an important role in this process, since they are very efficient in collecting solid particles in their center \citep{barge}. It has been proposed that this can speed up the formation of protoplanets enormously \citep{lyra}. It has also been claimed that vortices can effectively transport angular momentum outward \citep{johnson05}, allowing for an inward accretion flow in disks that have too low ionization fractions to be unstable to the MRI  \citep[magneto-rotational instability,][]{balbus}. 

The emergence and subsequent survival of vortices in protoplanetary disks has received a lot of attention recently. They can appear due to Rossby wave instabilities \citep{lovelace}, edge instabilities of planetary gaps \citep{koller,comparison,devalborro,minkai}, in 3D circulation models \citep{barranco} and in MHD turbulence \citep{fromang}. Vortex generation due to a baroclinic instability in disks with a radial entropy gradient was first proposed in \cite{klahr03}. However, this study was hampered by numerical issues, and could subsequently not be reproduced in a local shearing box \citep{johnson}. \cite{petersena} pointed out the importance of thermal diffusion in this problem, and showed, using anelastic global simulations, that vortices can be produced and maintained \citep{petersenb} in disks with a negative radial entropy gradient. More recently, \cite{lesur10} showed that the baroclinic instability is in fact subcritical and needs a finite amplitude disturbance to get going. They also showed that the vortices produced from this Subcritical Baroclinic Instability (SBI) can survive in 3D, despite the fact that they are unstable to parametric instabilities \citep{lesur09}.  

In this paper, we take the emergence of vortices to be given, and focus on one aspect of their subsequent evolution that has not been studied so far: orbital migration. It is well known that solid bodies embedded in Keplerian gas disks change their orbit through interaction with the gas. Small bodies experience a head wind from the gas, which is partly pressure supported, causing them to lose angular momentum and spiral inward \citep{weidenschilling}. The gravitational perturbation of a planet embedded in the disk leads to wave excitation at Lindblad resonances \citep{gt79}, which is asymmetric in general \citep{ward}, leading to inward migration for reasonable disk parameters \citep{tanaka}. This is called Type I migration. Interestingly, vortices are known to excite density waves in compressible simulations \citep[see][]{lesur10,tobyI,tobyII}, and one might expect that the same geometric effects that cause the torque imbalance in planetary Type I migration could lead to radial migration of vortices. This is the subject of this paper. 
 
The plan of the paper is as follows. We introduce the disk models in Sect. \ref{secEq}, and discuss the numerical methods used in Sect. \ref{secNum}. We then look at a particular example of vortex migration in Sect. \ref{secEx}, showing that this migration is due to the emission of density waves into the disk. We discuss the various physical mechanisms that lead to asymmetric wave emission in Sect. \ref{secWave}, show that this leads to vortex migration in Sect. \ref{secAngMom}, and compare these findings to numerical results in Sect. \ref{secGlob}. We discuss our results in Sect. \ref{secDisc}.

\section{Basic equations and disk models}
\label{secEq}
We consider thin disks in the two-dimensional approximation, and work with vertically averaged quantities only. The governing equations are then the continuity equation
\begin{equation}
\frac{\partial \Sigma}{\partial t}+\nabla \cdot \Sigma{\bf v}=0,
\label{eqCont}
\end{equation}
where $\Sigma$ is the surface density and ${\bf v}=(v_r,r\Omega)^T$ is the two-dimensional velocity in cylindrical coordinates (here $T$ indicates transpose), and the momentum equation
\begin{equation}
\frac{\partial {\bf v}}{\partial t}+{\bf v}\cdot \nabla {\bf v}=-\frac{\nabla p}{\Sigma}-\nabla \Phi,
\label{eqMom}
\end{equation}
with $p$ the vertically integrated pressure and $\Phi$ the gravitational potential. We consider inviscid disks only, and ignore the self-gravity of the gas, so that $\Phi$ is due to the central star only.  We will consider a strictly isothermal equation of state, $p=c_s^2 \Sigma$, with $c_s$ the sound speed. 

The disk surface density profile is a power law initially, $\Sigma=\Sigma_0 (r/r_0)^{-\alpha}$, with $r_0$ a reference radius, most often the initial location of the vortex. Since we do not consider self-gravity, $\Sigma_0$ is arbitrary. The sound speed is set by choosing a scale height $H$ at $r_0$, $H_0$, which determines the sound speed $c_s=H_0\Omega_0$. We then have $H=H_0(r/r_0)^{3/2}$. The angular velocity $\Omega$ is Keplerian, initially, with a small correction due to a radial pressure gradient. In some of the analysis, it is advantageous to have a constant aspect ratio $H/r$. This can be achieved by having a radially varying, but constant in time, sound speed. Physically, this corresponds to a disk that relaxes to an equilibrium temperature profile on a very short time scale (the cooling time scale is formally zero, so no temperature fluctuations are allowed).

It is possible to combine equations (\ref{eqCont}) and (\ref{eqMom}) into a single equation for the vortensity $\omega/\Sigma$, where $\omega={\bf \hat k} \cdot (\nabla \times {\bf v})$:
\begin{equation}
\frac{\partial}{\partial t} \left(\frac{\omega}{\Sigma}\right)+{\bf v}\cdot \nabla \left(\frac{\omega}{\Sigma}\right) = \frac{\nabla \Sigma \times \nabla p}{\Sigma^3},
\end{equation}
showing that in 2D barotropic flow, for which $p=p(\Sigma)$, vortensity is conserved along streamlines. This of course includes the isothermal case. It is no longer true when the sound speed varies radially, which makes the flow non-barotropic, or when 3D motions are allowed for. In that case, a more general quantity $(\nabla \times {\bf v})\cdot \nabla Q/\rho$, with $Q$ any quantity that is conserved on fluid elements and $\rho$ the density, is conserved along streamlines. For example, for an adiabatic flow $Q$ can be taken to be the specific entropy. The resulting quantity is sometimes called the potential vorticity.

On top of the equilibrium configuration we introduce a vortical perturbation. In global models, this is done by applying a circular velocity perturbation with a specified maximum over a circular patch of the disk. The velocity perturbation is usually a sizeable fraction ($\sim 0.5$) of the local sound speed, and the patch a sizeable fraction of the local scale height, typically $H_0/2$. The velocity perturbation tapers off exponentially away from the maximum. The exact form of the perturbation was found not to be important, as the system relaxes to a similar vortensity profile on a dynamical time scale regardless. Note that the subcritical baroclinic instability will tend to produce vortices of a size comparable to $H$ \citep{lesur10}. 

\section{Numerical methods}
\label{secNum}

For our global disk models, we solve equations (\ref{eqCont}) and (\ref{eqMom}) in cylindrical geometry $(r,\varphi)$ using {\sc rodeo} \citep{rodeo}, which is a second-order finite volume method using an approximate Riemann solver due to \cite{roe}. It has been successfully applied to the study of disk-embedded planets \citep{comparison,paardpap08} and disks in tight binary systems \citep{binary}. 

The computational domain consists of an annulus at a reference radius $r_0$, with inner and outer boundaries located at $0.2$ $r_0$ and $2.5$ $r_0$, respectively. We use non-reflective boundary conditions \citep{godon}, to allow the waves excited by the vortex to leave the computational domain freely. Although designed to be non-reflective for 1D simple waves only, we have observed no wave reflection in our 2D non-linear simulations. As explained in more detail in \cite{rodeo}, the use of a characteristics-based Riemann solver ensures that information is always drawn from the right places. At the boundary, any incoming characteristics are treated as if they originate from an unperturbed disk. This ensures that no waves enter the computational domain. We consider the full $2\pi$ in azimuth. For $H_0=0.1 r_0$, we use a typical grid size of 1536 in the radial and 6144 in the azimuthal direction, varying this up and down by factors of 2 to study effects of resolution. Our typical grid therefore has 67 zones per scale height. When considering smaller values of $H_0$, we increase the resolution to maintain a fixed number of grid cells per scale height. Within the code, we set $r_0=1$ for numerical convenience.  

\section{A fiducial case}
\label{secEx}

\subsection{Orbital migration}

\begin{figure}
\centering
\resizebox{0.8\hsize}{!}{\includegraphics[]{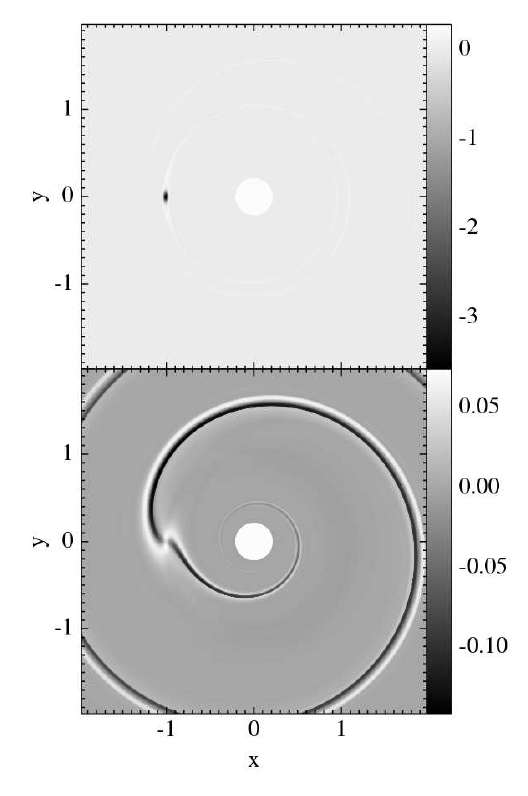}}
\caption{Relative perturbation of vortensity (top) and density (bottom) after 10 orbits at $r=1$ for an isothermal disk with $H_0=0.1 r_0$, $\alpha=3/2$, and an initial velocity perturbation of $0.5 c_s$ over a circular region of radius $H_0/2$ around $r=1$,$\varphi=\pi$.}
\label{figvortdens}
\end{figure}

In this section, we consider an isothermal disk with $\alpha=3/2$ and $H_0=0.1r_0$. Note that this is a disk with constant vortensity, and that, for numerical convenience, it is relatively thick for a protoplanetary disk. The dependence on $H_0$ is discussed in Sect. \ref{secGlob}. We give the disk a circular velocity perturbation of magnitude $c_{s}/2$ over a scale $H_0/2$. The resulting vortensity and density perturbations after 10 orbits at $r=1$ are shown in Fig. \ref{figvortdens}.

From Fig. \ref{figvortdens} we see that the vortensity perturbation remains localized, and that its radial extent roughly corresponds to the size of the initial perturbation. The density perturbation inside the vortex is modest ($\sim 5 \%$) compared to the vortensity perturbation ($\sim 300\%$), which indicates that on scales $\ll H$, the disk response to the perturbation is approximately incompressible, as expected. For a vortex in equilibrium in a shear flow, there exists a relation between the vorticity perturbation and aspect ratio \citep{kida81,lesur09}. In equilibrium, the relative perturbation in vorticity in a Keplerian disk should be $(\omega-\omega_0)/\omega_0=-3(\chi+1)/(\chi^2-\chi)$, where $\chi$ is the aspect ratio of the vortex. The strong vortex depicted in Fig. \ref{figvortdens} has an aspect ratio of $\sim 2.5$, which would lead to an equilibrium relative vorticity perturbation of $\sim -3$. Noting that we can neglect density perturbations in the almost incompressible disk response, and can therefore equate vorticity and vortensity relative perturbations, this is in good agreement with the top panel of Fig. \ref{figvortdens}. We note that this equilibrium is established for a range of vortex aspect ratios within a few orbits of the vortex. We therefore conclude that the vortices quickly adopt an equilibrium configuration, regardless of the initial perturbation.

\begin{figure}
\centering
\resizebox{0.8\hsize}{!}{\includegraphics[]{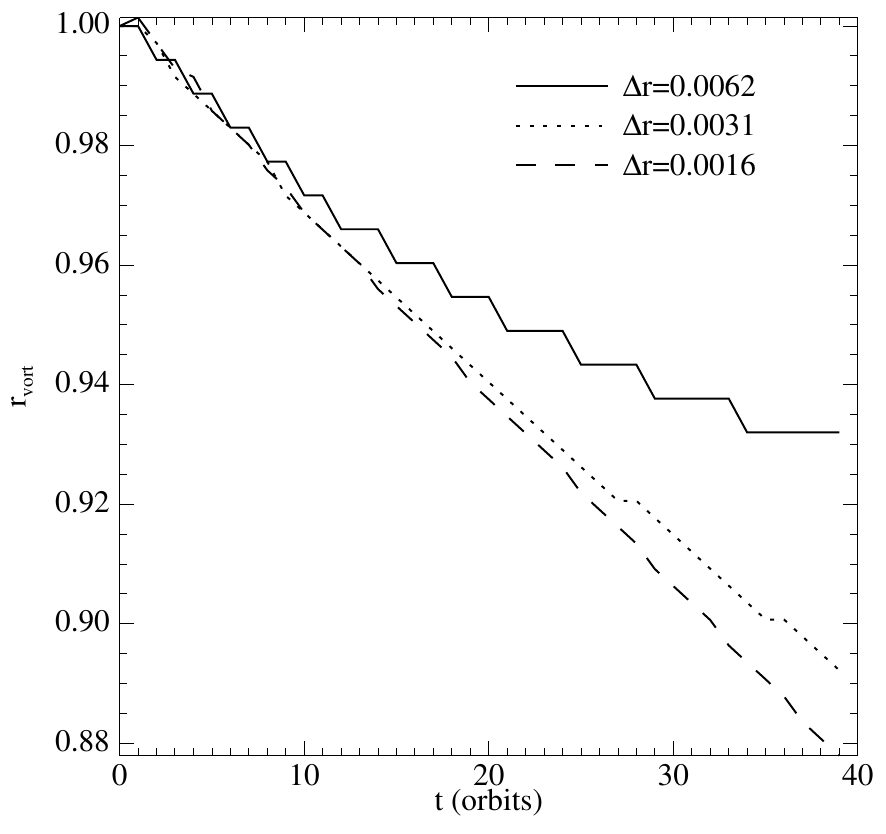}}
\caption{Time evolution of the radial location of the vortex, for three different resolutions. Disc and vortex parameters are the same as in Fig. \ref{figvortdens}.}
\label{figvortmig}
\end{figure}

The most notable density features start at a radial distance of approximately $H$ from the vortex, where the flow relative to the vortex becomes supersonic and density waves can be excited. The resulting two-armed spiral wave pattern is reminiscent of an embedded planet \citep{bryden}, with one inner wave propagating into the inner disk, and one outer wave propagating into the outer disk. For embedded planets, it is well-known that geometrical effects tend to favor the outer wave, which removes angular momentum from the planet and causes inward migration \citep{tanaka}. An additional complication in the vortex case is that the vortex may be asymmetric with respect to $r_0$. We deal with vortex asymmetries in Sect. \ref{VORTASM}; here, we just note that for $\alpha=3/2$, the vortex is symmetric to a high degree. The resulting migration of the vortex is shown in Fig. \ref{figvortmig}.  

Indeed the vortex migrates inward. Results are shown for three resolutions in Fig. \ref{figvortmig}. The lowest resolution has 16 cells per scale height, radially, at $r_0$. In other words, the vortex is resolved by approximately 8 cells only, in the radial direction. In this case, the vortex weakens through numerical diffusion, which starts to slow down the migration rate after 10 orbits. Doubling the resolution pushes this time towards 20 orbits, and doubling it again gives a steady migration rate for at least 40 orbits. At early times, the migration rate is similar for all resolutions, and we conclude that this migration rate is converged with respect to numerical resolution. It results in a migration time scale of $r_0/|\dot r| \approx 2000~ \Omega_0^{-1}$, or $300$ orbits at $r=r_0$. It is clear that for this disk, vortex migration is an important process to consider. Additional runs that included viscosity, with the kinematic viscosity $\nu$ parametrised using the $\alpha$-prescription, $\nu=\alpha_v c_s H$, indicated that the numerical dissipation of these vortices roughly correspond to $\alpha_v=10^{-4}$, $10^{-5}$ and $10^{-6}$ respectively for the low, medium and high resolution cases shown in Fig. \ref{figvortmig}.

\subsection {Wave action}
Density waves propagate inwards and outwards away from the  vortex. Provided the vortex is not too strong, we expect the waves to be in the linear regime close to the vortex, becoming non linear at larger distances where shocks are formed \citep[e.g.][]{goodraf} and their amplitude reduces. While in the linear regime, the total rate of flow of angular momentum across a radial location that is  associated with the waves, or equivalently the wave action, is conserved. When the non linear regime is entered and dissipation occurs, the wave action decreases. Angular momentum  is advected   inwards in the  radial direction at a rate given by
\begin{equation}
A=2\pi r^2\langle \Sigma v_r v_{\varphi}\rangle ,\label{Waction}
\end{equation}
where the angle brackets denote an azimuthal average. Because there is no net mass flow associated with linear waves, the azimuthal velocity perturbation alone may be used in (\ref{Waction}) in this case. We remark that, by making use of the linearized equations governing the waves  is is possible to write (\ref{Waction}) in a different form. The waves have a pattern speed $\Omega_\mathrm{vort}$  which is  also the angular velocity of the vortex. Thus for the perturbations associated with them we have
\begin{equation}
\Omega_\mathrm{vort} \frac{\partial}{\partial \varphi}=\frac{\partial}{\partial t}.
\end{equation}
Accordingly the perturbed azimuthal component of the equation of motion gives
\begin{equation}
(\Omega-\Omega_\mathrm{vort} )\frac{\partial v_{\varphi}}{\partial \varphi}+\frac{\kappa^2}{2\Omega}v_r =
-\frac{c_s^2}{r\Sigma}\frac{\partial \Sigma'}{\partial \varphi},
\end{equation}
where $\Sigma'$ is the surface density perturbation. Noting that to linear order,  the Lagrangian displacement $\xi_r$ satisfies
\begin{equation}
v_r =\left(\Omega- \Omega_\mathrm{vort}\right ) \frac{\partial \xi_r }{\partial \varphi}\label{Lagdisp},
\end{equation}
we see that
\begin{equation}
\left( \Omega-\Omega_\mathrm{vort} \right )\left( v_{\varphi}+\frac{\kappa^2}{2\Omega}\xi_r\right) =
-\frac{c_s^2\Sigma'}{r\Sigma}.\label{wvaction}
\end{equation}
Using (\ref{wvaction}) to  substitute for  $v_{\varphi}$ in (\ref{Waction}) and making use of (\ref{Lagdisp}), it is readily seen that we can rewrite (\ref{Waction}) in the form
\begin{equation} 
A= 2\pi \left \langle \frac{r c_s^2 \Sigma'v_r}{\Omega_\mathrm{vort}-\Omega}\right\rangle.
\label{eqwaveaction} 
\end{equation} 

We can now verify that this migration is due to the density waves by studying  the behavior of  wave action. The change in the angular momentum content of the fluid on account of the presence
of the vortex is
\begin{equation}
L_\mathrm{vort}=\int (\Sigma r^2\Omega-\Sigma_i r^2\Omega_i) dS,
\end{equation}
where subscripts $i$ denote the initial conditions, and the integral is taken over the whole vortex surface area. We expect $dL_\mathrm{vort}/dt=\Delta A$, with $\Delta A$ the difference in wave action defined above evaluated for the outer  and the inner wave. 

\begin{figure}
\centering
\resizebox{0.8\hsize}{!}{\includegraphics[]{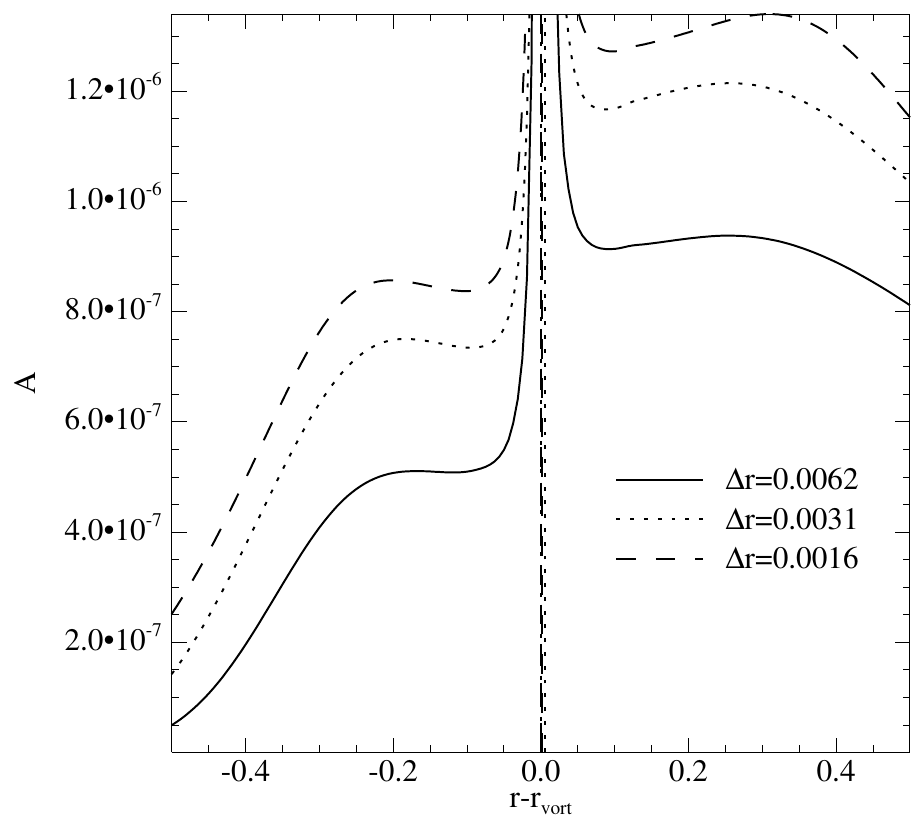}}
\caption{Wave action after 10 orbits for the same vortices as in Fig. \ref{figvortmig}.}
\label{figwaveaction}
\end{figure}

We show the wave action for three resolutions in Fig. \ref{figwaveaction}. Equation \ref{eqwaveaction} only has meaning in the wave region, at distances of at least $2H/3$ from the vortex. Although our lowest resolution run clearly does not resolve the waves well enough to get the correct wave action on both sides, $\Delta A$ is very similar to the higher resolution runs. In Sect. \ref{secGlob}, we make sure our resolution is at least as high as that of the dashed line in Fig. \ref{figwaveaction}. The similarity of the wave action for the highest resolution runs translates into a similar migration rate after 10 orbits (see Fig. \ref{figvortmig}). We have measured the change in angular momentum of the vortex between 5 and 10 orbits, and compared this to what is expected from the wave action. The results are summarized in Table \ref{tabwave}.

\begin{table}
\caption{Vortex angular momentum and wave action}
\begin{tabular}{ccc}
\hline
$\Delta r\footnote{Change in orbital distance of the vortex between $5$ and $10$ orbits.}$ & $\Delta A\Delta t\footnote{Angular momentum transport by density waves.}$ & $\Delta L\footnote{Change in angular momentum of the vortex between $5$ and $10$ orbits.}$ \\
\hline
0.0062 & $-1.1\cdot 10^{-5}$ & $-7.3\cdot 10^{-6}$\\
0.0031 & $-1.4\cdot 10^{-5}$ & $-6.7\cdot 10^{-6}$ \\
0.0016 & $-1.6\cdot 10^{-5}$ & $-1.3\cdot 10^{-5}$ \\
\hline
\end{tabular}
\label{tabwave}
\end{table}

We point out that there is a large uncertainty in the measured $\Delta L$. This is because $L_\mathrm{vort}$ can only be measured to an accuracy of $\sim 5\%$ because of ambiguities in defining the vortex as a separate entity from the background flow, and $\Delta L$ results from a subtraction of two almost equal numbers (typically $\Delta L \approx 0.1L_\mathrm{vort}$). We therefore estimate that there is an uncertainty of at least $50\%$ in the measured values of $\Delta L$, which makes the agreement between $\Delta A\Delta t$ and $\Delta L$ reasonable (see Table \ref{tabwave}). Therefore, the inward migration of the vortex is consistent with the wave torque.   

\begin{figure}
\centering
\resizebox{0.8\hsize}{!}{\includegraphics[]{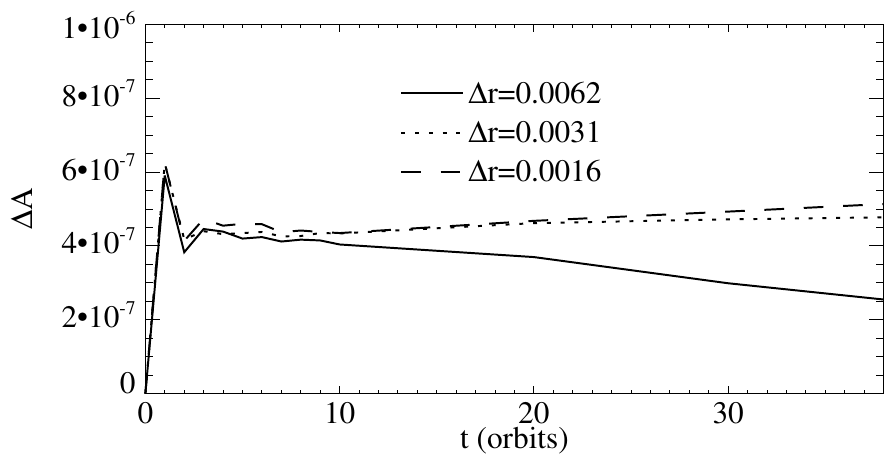}}
\caption{Time evolution of the wave action difference, for the same three resolutions as in Fig. \ref{figvortmig}.}
\label{figdAdt}
\end{figure}

We show the time evolution of the wave action difference, or wave torque, in Fig. \ref{figdAdt}, for the same three resolutions as in Fig. \ref{figvortmig}. In all cases, a steady torque is set up after 3 orbits, at which time the vortex has reached its equilibrium with the background shear. After the equilibrium has been set up, the slow time evolution of the torque is due to numerical diffusion (the decline in the low resolution case) and slowly varying background disk properties due to the changing radial location of the vortex (the slow increase in the highest resolution cases).

\section{Wave asymmetries}
\label{secWave}

\subsection{Linear equations}

In order to understand the origin of the waves emitted by accretion disk vortices, let us introduce the local shearing box model \citep{hawley95}. In this model, we neglect curvature effects in equations (\ref{eqCont})-(\ref{eqMom}) and introduce a local cartesian frame corotating with the disk at $r_0$ by defining $x=r-r_0$ and $y=r_0\varphi$. In this model, equations (\ref{eqCont}) and (\ref{eqMom}) read:
\begin{eqnarray}
\label{eqContLoc}\frac{\partial \Sigma}{\partial t}+\bm{\nabla \cdot} \Sigma\bm{v}=&0,\\
\label{eqMomLoc}\frac{\partial \bm{v}}{\partial t}+\bm{v\cdot \nabla v}=& \nonumber \\
-c_s^2\bm{\nabla} \ln\Sigma-\bm{\nabla} (-q\Omega^2x^2)-2\bm{\Omega \times v},&
\end{eqnarray}
where $\Omega=\Omega(r_0)$ is the Keplerian frequency at $r_0$, $q=3/2$ for a Keplerian rotation profile and we have assumed an isothermal gas. In the following, we consider a vortex located at $r=r_0$ and perturbing the surrounding flow. The vortex core is a non-linear solution to the above equations (see e.g. \citealt{lesur09}), and we consider the \emph{linear} perturbations produced by the vortex core at long distance. Since the vortex is, in first approximation, a quasi-steady structure, we consider stationary perturbations of the Keplerian flow $\bm{v}_0=-q\Omega\bm{e_y}$ introducing
\begin{eqnarray}
\Sigma'(x,y)&=&\Sigma(x,y)-\Sigma_0,\\
\bm{u}(x,y)&=&\bm{v}-\bm{v}_0.
\end{eqnarray}
As a further simplification, we only consider the case of a constant background density profile $\Sigma_0$. To reduce the problem to a system of ordinary differential equations, we introduce a Fourier decomposition in the $y$ direction:
\begin{eqnarray}
\Sigma'&=&\tilde{\Sigma}'(x)\exp (iky),\\
\bm{u}&=&\tilde{\bm{u}}(x)\exp (iky).
\end{eqnarray}
Using this decomposition in (\ref{eqContLoc}) and (\ref{eqMomLoc}) yields:
\begin{eqnarray}
\label{eqLinCont}-i\sigma(x)\tilde{\Sigma}'+\Sigma_0\Big (\frac{d \tilde{u}_x}{dx}+ik\tilde{u}_y\Big )&=&0\label{LE1}\\
\label{eqLinMomx}-i\sigma(x)\tilde{u}_x-2\Omega \tilde{u}_y+\frac{c_s^2}{\Sigma_0}\frac{d\tilde{\Sigma}'}{dx}&=&0\label{LE2}\\
\label{eqLinMomy}-i\sigma(x)\tilde{u}_y+(2-q)\Omega \tilde{u}_y+ic_s^2k\frac{\tilde{\Sigma}'}{\Sigma_0}&=&0
\end{eqnarray}
where we have defined $\sigma(x)=q\Omega x$. After some algebra, one gets the following equation for the azimuthal velocity $\tilde{u}_y$:
\begin{equation}
\label{eqParabol}\frac{d^2\tilde{u}_y}{dx^2}+\Lambda^2(x)\tilde{u}_y=0,
\end{equation}
where 
\begin{equation}
\Lambda^2(x)=\frac{\sigma^2(x)-2(2-q)\Omega^2}{c_s^2}-k^2.
\end{equation}
This equation describes the propagation of a linear acoustic-inertial wave in the shearing box. As stated above, we consider a nonlinear vortex core located at $x=0$ emitting a wave at large distance. Assuming the radial extent of the vortex is very small compared to the disk thickness $H=c_s/\Omega,$ although this need not be the case for the azimuthal extent,  the vortex core can be seen in first approximation as a boundary condition at $x=0$ for the wave emission problem. Without loss of generality, we will consider the case of a wave emitted in the $x>0$ region, the $x<0$ solutions being deduced by symmetry arguments.

It is worth noting that (\ref{eqParabol}) is in fact the vorticity conservation equation in a compressible medium, with a constant background vorticity $(2-q)\Omega$. In a global disk however, background vorticity is not constant and density might also depend on radius. In this context, one would write a conservation equation for the vortensity $\bm{\nabla \times v}/\Sigma$ which would be similar to (\ref{eqParabol}) if the background vortensity was constant. Therefore, the wave emission model we are discussing here should be understood as a local model for a \emph{constant vortensity} disk.

\subsection{Wave solutions}
\begin{figure}
\centering
\resizebox{0.9\hsize}{!}{\includegraphics[]{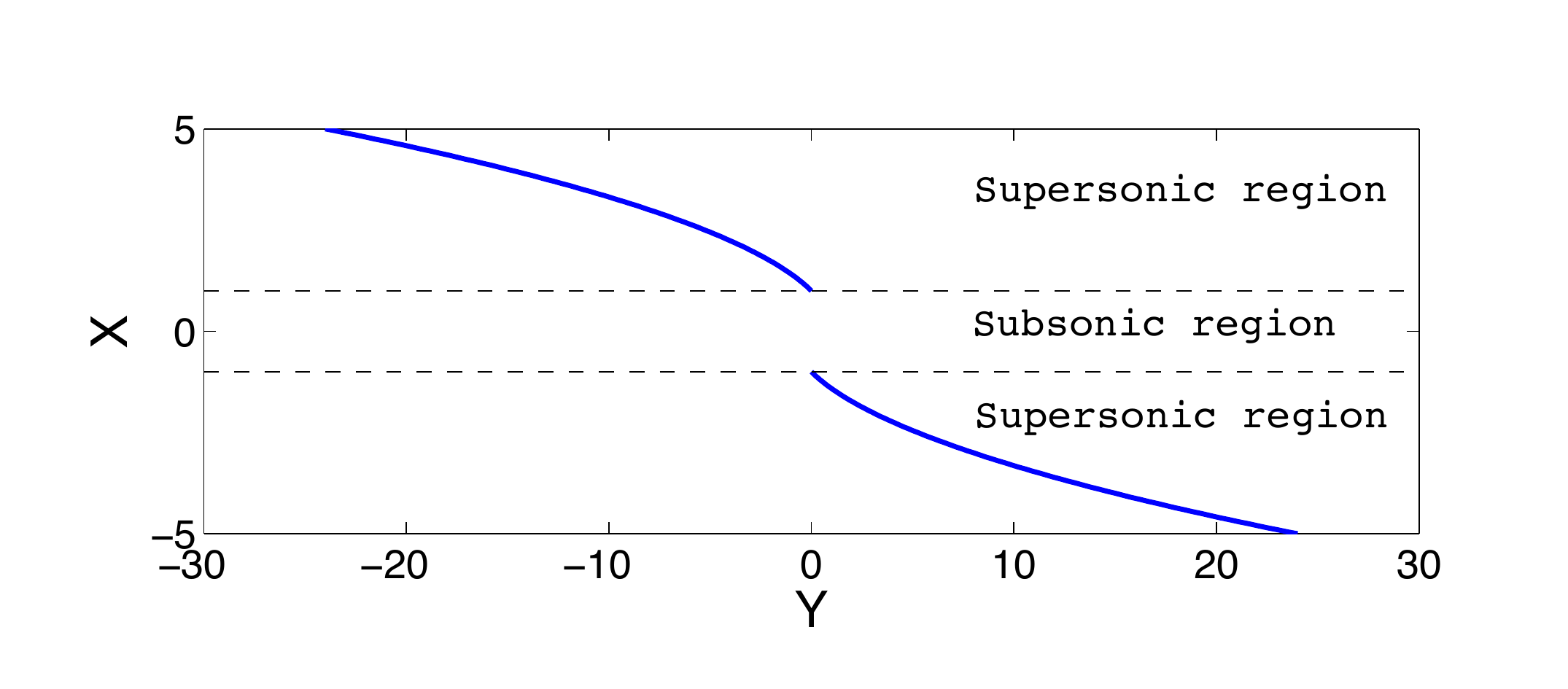}}
\caption{Supersonic solution wave fronts as obtained from (\ref{eqWaveFront}) assuming $x_s=1$. The sonic lines are indicated by dashed lines.}
\label{figWakes}
\end{figure}

The above equation admit solutions in the form of parabolic cylinder functions. However, to understand the physical origin of vortex waves, we present here WKB solutions. We first note that (\ref{eqParabol}) has a turning point at 
\begin{equation}
\nonumber x_s=\frac{\omega_{ai}}{q\Omega k}.
\end{equation}
where we have introduced $\omega_{ai}=\sqrt{k^2c_s^2+q(2-q)\Omega^2}$, the natural frequency of acoustic-inertial waves. This particular location corresponds to the sonic line of the vortex. This is the line along which the flow speed is equal to the wave speed  as given by the phase velocity. In the limit of large  $k$  this phase velocity becomes equal to the sound speed hence the term  sonic line. For $x<x_s$ (``subsonic'' region), the wave has an exponential shape and it corresponds to the logarithmic tail of an incompressible vortex. Using first order WKB theory in the limit $x\ll x_s$, we get:
\begin{eqnarray}
\tilde{u}_y=A\exp\Bigg(-\int_x^{x_s} \Big[-\Lambda^2(u)\Big]^{1/2}\,du\Bigg)+\nonumber \\
B\exp\Bigg(\int_x^{x_s}  \Big[-\Lambda^2(u)\Big]^{1/2}\,du\Bigg).
\end{eqnarray}
On the other hand, for $x>x_s$ (``supersonic'' region), the wave is essentially sinusoidal and can travel for very large distances. In the limit $x\gg x_s$, we obtain:
\begin{eqnarray}
\tilde{u}_y=C\exp\Bigg(i \int_x^{x_s} \Big[\Lambda^2(u)\Big]^{1/2}\,du\Bigg)+\nonumber \\
D\exp\Bigg(-i \int_x^{x_s}  \Big[\Lambda^2(u)\Big]^{1/2}\,du\Bigg),
\end{eqnarray}
where it can be shown that an outgoing wave (wave transporting angular momentum outward) is obtained when $C=0$. This solution leads to the long ``wakes'' one observes in compressible simulations of vortices. In particular, in the limit $x\gg kc_s/q\Omega$, we find that the wavefronts of the supersonic region are given by
\begin{equation}
\label{eqWaveFront}y=y_0-\frac{q\Omega}{2 c_s}x^2,
\end{equation}
which produce a pattern similar to the wakes found in simulations (Fig.~\ref{figWakes}).

The connection of the supersonic and subsonic regions allows one to determine $C$ and $D$ knowing $A$ and $B$. This can be done using asymptotic matching techniques based on Airy functions. Although the precise derivation of these solutions is beyond the scope of this paper, we find that $|C|\sim|D|\sim {\rm max}(|A|,|B|)$, i.e. the amplitude of the solution does not change significantly as one crosses the sonic line. However, it should be noted that the wave amplitude in the supersonic region is significantly reduced compared to its amplitude at $x=0$ due to the exponential decay in the subsonic region. One can estimate this attenuation factor $\Gamma$ with:
\begin{eqnarray}
\Gamma^\pm=\frac{|\tilde{u}_y(\pm x_s)|}{|\tilde{u}_y(x_0)|}&\simeq&\exp\Bigg(\mp \int_0^{\pm x_s}  \Big[-\Lambda^2(u)\Big]^{1/2}\,du\Bigg)\nonumber \\
&=&\exp\Big(-\frac{\pi\omega_{ai}^2}{4q\Omega kc_s}\Big ).
\label{eqattenuation}
\end{eqnarray}
This attenuation factor has a simple physical interpretation: it relates the vortex amplitude to the emitted density wave amplitude at a particular wavenumber. One can show that the attenuation factor is minimum for $k=\sqrt{q(2-q)} H^{-1}$ where $H$ is the disk thickness $H\equiv c_s/\Omega$. 

It is clear that in the shearing-box the wave emitted in the inner side and the wave emitted in the outer side have the same amplitude ($\Gamma^+=\Gamma^-$). Because of this symmetry, one does not expect any migration. However, if one includes global effects such as curvature and the dependence of $\Omega$ as a function of radius, then an asymmetries might occur, resulting in vortex migration.

\subsection{Cylindrical Geometry}
\label{secCyl}
To extend our analysis to cylindrical geometry, we consider the case of a constant vortensity disk $\Sigma\propto r^{-3/2}$ with a constant aspect ratio $c_s\propto r^{-1/2}$. Furthermore, we assume the flow is barotropic and work in an inertial frame. We Fourier transform the velocity field in $t$ and $\varphi$, defining $\bm{u}(t,\varphi,r)= \sum_m \int d\omega \exp[ i(\omega t-m\varphi)] \tilde{\bm{u}}(\omega,m,r)$. For a vortex corotating with the gas at $r=r_0$ we adopt  $\omega =m\Omega(r_0).$ We then  perform the corresponding  linearization  of the equations  to that described in the shearing box case. One eventually obtains:
\begin{equation}
\frac{1}{\Omega r}\frac{d}{dr}\Omega r \frac{d}{dr} r\tilde{u}_\varphi+\Lambda^2(r) r\tilde{u}_\varphi=0
\end{equation}
where
\begin{equation}
\Lambda^2(r)=\frac{\sigma^2(r)-\Omega^2}{c_s^2}-\frac{m^2}{r^2}
\end{equation}
and $\sigma(r)=\omega-m\Omega(r)$. This equation is very similar to the shearing-sheet problem. As in the shearing sheet case, we find two turning points $r_s^\pm$ on both sides of the corotation radius $r_0$ defined as $\sigma(r_0)=0$. However, it should be noted that $|r_s^+-r_0| < |r_s^--r_0|$, i.e. the outer sonic line is closer to corotation than the inner sonic line. Indeed, imposing $\Lambda(r_s^\pm)=0$ leads to:
\begin{equation}
r_s^{\pm}=r_0\Bigg(1\pm\sqrt{\frac{1}{m^2}+\epsilon^2}\Bigg)^{2/3}\label{sonlin}
\end{equation}
where $\epsilon$ is the constant aspect ratio $\epsilon=H/r$.
In the limit $m\rightarrow \infty$ we have
\begin{equation}
r_s^{\pm}=r_0\Bigg(1\pm\frac{c_s}{r\Omega}\Bigg)^{2/3}.
\end{equation}
Hence in this limit the sonic lines occur where the  velocity  relative to the vortex, that is associated with the background shear, is equal to  the sound velocity. This corresponds to the large $k$ limit in the local model. However, this does not occur for small $m.$ In fact (\ref{sonlin}) indicates that the distance  of the sonic lines from the vortex increases as $m$ decreases to the extent that when $m=1,$ the inner sonic line is out of the domain. This is because the phase velocity of the density waves increases as $m$ decreases. WKB solutions very similar to the one presented for the shearing box case can be derived and one can compute the attenuation factor as:
\begin{equation}
\Gamma^\pm=\frac{r_s^\pm|u_\varphi(r_s^\pm)|}{r_0|u_\varphi( r_0)|}\simeq\exp\Bigg(\mp \int_{r_0}^{r_s^\pm}  \Big[-\Lambda^2(u)\Big]^{1/2}\,du\Bigg).
\end{equation}
As it can be seen from this expression, the attenuation factor on the inner side ($\Gamma^-$) is expected to be different from the attenuation factor on the outer side $\Gamma^+$ because $\Lambda$ is not symmetric on both sides of corotation. One can easily check that $|\Lambda^2(r_0+\delta r)|<|\Lambda^2(r_0-\delta r)|$ when $\delta r < r_s^+-r_0$. Moreover, since $|r_s^+-r_0| < |r_s^--r_0|$, we conclude that $\Gamma^+>\Gamma^-$, i.e. the outer wave always have an amplitude larger than the inner wave. 

This can also be seen by making the transformation from $r$ to $\psi$ such that
\begin{equation}
\frac{r}{r_0}=1+\sqrt{\epsilon^2+\frac{1}{m^2}}\cos\psi,
\end{equation}
so that the domain $(r_s^+,r_0)$ is mapped to $(0,\pi/2)$ and the domain $(r_s^-,r_0)$ is mapped to $(\pi/2,\pi).$ The integrals
\begin{equation}
\ln \Gamma^\pm = \mp\int_{r_0}^{r_s^\pm}  \Big[-\Lambda^2(u)\Big]^{1/2}\,du
\end{equation}
transform to
\begin{equation}
\ln \Gamma^\pm = -\int_{-\pi(-1\pm 1)/4}^{-\pi(-3\pm 1)/4} \left(\frac{2m}{3\epsilon}\right) 
\frac{(\epsilon^2+\frac{1}{m^2})\sin^2\psi}{ 1+ \sqrt{\epsilon^2+\frac{1}{m^2}}\cos\psi}d\psi.
\end{equation}
On account of the differing integration ranges for the outward and inward propagating waves the wave attenuation factor  is readily seen to   be larger for the inward propagating wave. Although the calculation can be done in full analytically, it is adequate for our purposes to regard $\epsilon$ and $1/m$ to be comparable small parameters and express the result correct up to order $\epsilon.$ Then
\begin{equation} 
\ln \Gamma^\pm = - \frac{\pi }{6}
\left(m\epsilon + \frac{1}{ m\epsilon}\right)\left(1\mp \frac{4}{3\pi}\sqrt{\epsilon^2+\frac{1}{m^2}}\right).
\end{equation} 
Thus the expected ratio of outward to inward wave flux arising from this asymmetry  is
\begin{equation}
\left(\frac{ \Gamma^+}{\Gamma^-}\right)^2  =\exp \left[- \frac{8 }{9}\sqrt{\frac{\epsilon}{m}}
\left(m\epsilon + \frac{1}{ m\epsilon}\right)^{3/2}\right]\label{WASM} .
\end{equation}
As we will see later, this explains the inward migration found in simulations with a similar profile. For example when $\epsilon =0.1$ and $m=5,$ (\ref{WASM}) predicts an asymmetry comparable to that seen in Fig. \ref{figwaveaction}. More generally, when   $m=1/\epsilon$ we obtain  a ratio $\sim 1 - 2.5/m.$
 
\subsection{Vortensity gradient}\label{VORTASM}

The above discussion on the local shearing box model assumed a constant vortensity and in this case the response to a symmetric vortex was found to be symmetric leading to equal inward and outward  angular momentum flow rates and hence no net migration. However, if the constraint of constant vortensity is relaxed an asymmetry in the response occurs in the subsonic region of the flow even when the vortex is symmetric. To see this we consider a local shearing box model in which the  background density  has the form $\Sigma={\rm const}\times \exp(-\alpha x)$ with $\alpha$ being a constant. The vortensity of course varies as the inverse of $\Sigma.$

As we are interested in the subsonic region  we assume the linear response satisfies  the anelastic condition $\nabla\cdot( \Sigma \tilde{{\bf  u}}) =0.$ After eliminating $\tilde{\Sigma}'$ from (\ref{LE1}) and (\ref{LE2}) and then  using the anelastic condition to eliminate $\tilde{u_y},$ we obtain an     equation for $Q=\tilde{v}_x\sqrt{\Sigma}$ in the form
\begin{equation}
\frac{d^2 Q}{dx^2}= Q\left(k^2-\frac{\alpha}{3x}\right),\label{VORTEQ}
\end{equation}
where  $k^2=m^2+\alpha^2/4.$ We remark that for assumed unit radius of the center of the box, $k_y \equiv m$ is the azimuthal mode number and $\alpha$ is the power law index occurring  in $\Sigma \propto r^{-\alpha}$ for a global disk.

We are interested in solutions of (\ref{VORTEQ}) that decay exponentially for large $|x|$ and are matched as $x \rightarrow 0.$ However, we do dot require that derivatives match for $x \rightarrow 0.$ This is because we assume that there is a nonlinear vortex core there that is not modeled by (\ref{VORTEQ}) which only describes the regions between the vortex core and the sonic lines. 

Without loss of generality we  shall consider  $x>0.$ The case $x<0$ can be recovered by letting $\alpha \rightarrow -\alpha.$ An appropriate integral representation for the required solution follows from representations of 'Coulomb Wave Functions' given in \cite{abrasteg} and may be written as
\begin{eqnarray}
 Q =C_0\Gamma(q)(2k)^{-q}\exp(-kx)& \nonumber \\ 
 \int ^{\infty}_0 \left(x+\frac{t}{2k}\right)^{-q}t^{q}&\exp(-t)dt.
\end{eqnarray}
where $C_0$ is an arbitrary  constant and $q=-\alpha/(6k).$
The above form may be used to connect the form of the  solution near $x=0$
to that for large $x$ giving
\begin{eqnarray}
&Q(x=0)=1 \rightarrow Q \nonumber \\
&\sim\Gamma(q+1)(2kx)^{-q}\exp(-kx)  \ \ \ \ {\rm for} \ \ \ \  x
\rightarrow \infty
\end{eqnarray}
 which may also be expressed as
\begin{eqnarray}
y(x=0)=1 \rightarrow \nonumber \\
y \sim \exp(-kx -q\ln (2kx) -\gamma q) \ \ \ \ {\rm for} \ \ \ \  x
\rightarrow \infty.
\end{eqnarray}
Here, assuming that $k$ is large,  we have used $\Gamma(q+1) \sim -\gamma q,$ where $\gamma$ is Euler's constant $\sim 0.577.$

Recalling  that $q = -\alpha/(6k),$ we find that the solution exceeds the pure exponential for large $x>0$ when $\alpha$ is positive. That means the outgoing wave is favored in this case. Conversely the ingoing wave is favored when $\alpha < 0$ indicating the potential for inward migration in this case. But, noting that $kx \sim 1$, this is found to lead only to a  relatively weak modification of the wave propagation, except when $|\alpha|$ is large and  the vortensity gradient is very strong.

\begin{figure}
\centering
\resizebox{0.80\hsize}{!}{\includegraphics[]{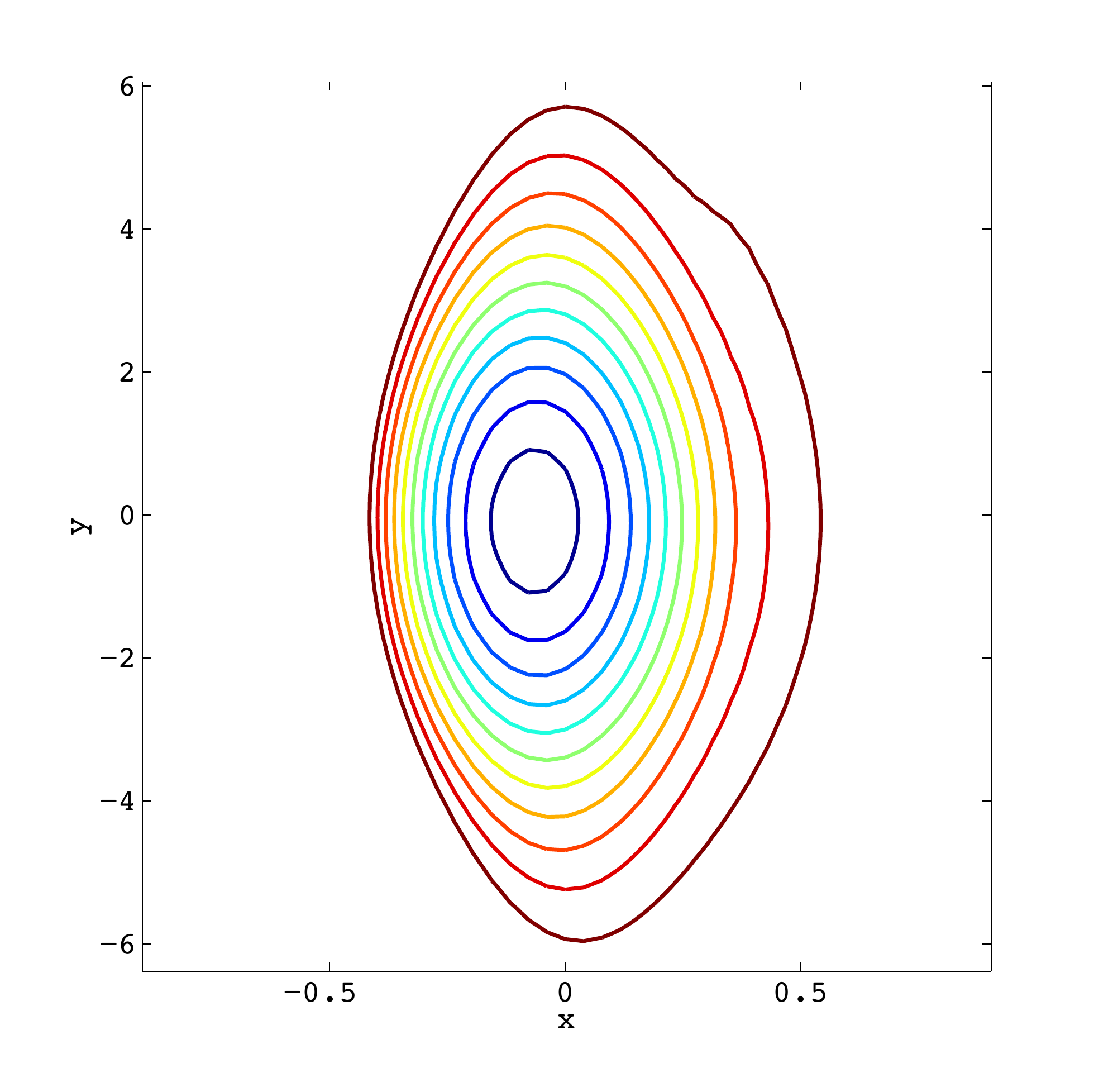}}
\caption{Isocontour of vorticity for a vortex embedded in a vortensity gradient. The vortex is deformed by the stratification and gets flattened on the denser region ($x<0$).}
\label{figAnelasticVortex}
\end{figure}

However, it should be noted that the presence of a vortensity gradient leads to the modification of the vortex structure \emph{itself}.  In the above calculation, we have assumed the vortex core was symmetrical, i.e. $u_x(x=+\epsilon)=u_x(x=-\epsilon)$ when $\epsilon\rightarrow 0$. However, in the presence of a vortensity gradient, vortex solutions are not symmetric anymore. To show this effect, we have  solved a set of anelastic equation in a shearing box including a radial vortensity gradient:
\begin{eqnarray}
\bm{\nabla \cdot}(\Sigma_0(x)\bm{v})&=&0,\\
\frac{\partial \bm{v}}{\partial t}+\bm{v\cdot \nabla v}&=&-\bm{\nabla} \Pi-\bm{\nabla} (-q\Omega^2x^2)-2\bm{\Omega \times v},
\end{eqnarray}
In these equations, the vortensity gradient is imposed by the density profile $\Sigma_0(x)$ and  density/sound waves are filtered out. We have integrated numerically these equations assuming an exponential profile for the surface density $\Sigma=\Sigma_0\exp(-\alpha x)$ using a spectral scheme. The initial conditions  were chosen to be those appropriate to  a Kida vortex \citep{kida81} of aspect ratio 8. This vortex  then relaxes to a new quasi-stationary state corresponding to a vortex solution with a vortensity gradient. We show in Fig.~\ref{figAnelasticVortex} the resulting vorticity distribution in the case $\alpha=4$ (the length unit being the vortex radial size). As it can be seen from this figure, the vortex core becomes asymmetric when a vortensity gradient is imposed. In particular, the perturbation  extends for a  greater distance on the less dense side and the vortex is ``flattened'' on the denser side. As a result, the less dense part of the vortex, being more extended,  will excite stronger density waves than the inner side, resulting in a new source of wave asymmetry. It should be noted that this effect is purely hydrodynamic and is due to the fact that the perturber (the vortex) ``feels'' and responds to the surrounding flow. Therefore, no corresponding effect exists in the case of a planet embedded in a disk.

\subsection{Comparison of vortex and planet migration}
\label{secCompare}

\begin{table}
\caption{Comparing planet and vortex migration}
\begin{tabular}{ccc}
\hline
Effect & Vortex & Planet \\
\hline
geometrical & yes & yes\\
core asymmetry & yes & no \\
corotation torque & no & yes \\
pressure buffer & no & yes\\
\hline
\end{tabular}
\label{tabmig}
\end{table}

We now summarize some important differences between vortex migration and Type I planet migration (see Table \ref{tabmig}). Although both are driven by asymmetric density waves, the excitation mechanism as well as the sources of the asymmetry are fundamentally different. First of all, it should be pointed out that although it has  very similar properties, in contrast to the planet migration problem, the asymmetry  discussed in Sect. \ref{secCyl} does not involve any torque calculation. Instead, it  appears as an attenuation in the linear propagation of density waves emitted by the vortex and it leads to a stronger outer wave. This is a major difference with the planet case in which asymmetric Lindblad torques  acting at the sonic line are believed to drive planet migration. Furthermore, since we find the vortex always to move with the local gas velocity, there is no effect of the pressure buffer as discussed in the appendix. We find that wave asymmetries only weakly depend on the background density profile, favoring the outer wave and therefore inward migration.

However, unlike a planet, a vortex directly feels the disk and its associated density profile, which leads to an asymmetric vortex core as discussed in Sect. \ref{VORTASM} in the case of a vortensity gradient. The resulting asymmetric wave emission can either counteract or reinforce inward migration due to geometrical effects as discussed above.  

We also point out that since there is no gravitational interaction between disk and vortex, there are no horseshoe orbits close to corotation. In the planet case, the horseshoe drag \citep{ward91} is an additional source of angular momentum exchange \citep{drag}, that can be strong enough to counteract the wave torque in non-isothermal disks \citep{paard10}. In the vortex case, we only have the wave torque.

\section{Angular momentum transport and vortex migration}
\label{secAngMom}

It can be shown that the density waves described above transport angular momentum outward \citep{papalin95}. Therefore, if the waves emitted by a vortex are asymmetric then angular momentum has to be deposited or extracted in the vortex neighborhood. This in turn can be linked to vortex migration since moving a vortex from one location to another means some angular momentum was exchanged between these two locations. This argument shows that one can in principle relate the vortex migration rate to the wave emission asymmetry, without knowing precisely how the waves interact with the vortex structure. To quantify this effect more precisely, let us start considering the angular momentum conservation equation:
\begin{equation}
\partial_t r \langle \Sigma (u_\varphi+r\Omega(r)\rangle+\frac{1}{r}\partial_r r \langle \mathcal{F}\rangle=0
\end{equation}
where $\mathcal{F}$ is the outward  angular momentum flux
 $\mathcal{F}=r\Sigma u_r (r\Omega(r)+u_\varphi)= A/(2\pi r)$ (see equation (\ref{Waction})) and $\langle \cdot \rangle$ denotes a $\varphi$ average.
Let us assume a vortex of size $2s$ is located at $r=r_0$ and emits density waves asymmetrically. Integrating the above equation between the inner and outer radius tells us the vortex is \emph{losing} angular momentum if $\mathcal{F}(r_0+s)>\mathcal{F}(r_0-s)$, i.e. if the outer wave is stronger than the inner wave, as in the constant vortensity disk described above. Next, to relate this effect to vortex migration, one has to compute the angular momentum of a vortex.

The angular momentum excess (compared to an unperturbed disk) due to a single vortex is given by:
\begin{equation}
\label{Jdef}
\mathcal{J}\simeq \int_{r_0-s}^{r_0+s} dr\,r^2(\delta \Sigma r\Omega(r)+\Sigma_0u_\varphi) \Delta\varphi
\end{equation}
where $\Sigma_0$ is the unperturbed density and the integration is computed between the inner and outer radius of the vortex. We denote the angular size of the vortex as $\Delta \varphi$, which we expect to be $\sim H_0/r_0$. It can be seen that the angular momentum of a vortex is made of two parts: a contribution due to the density fluctuation associated to the vortex, and a contribution due to the vortex rotation profile $u_\varphi$. Assuming the vortex rotation rate is $\omega$, the velocity can be approximated by $u_\varphi\simeq (r-r_0)\omega$ for $r_0-s<r<r_0+s$. The density fluctuation $\delta \Sigma$ can then be calculated assuming a geostrophic equilibrium in the vortex core. Using the  radial component of the equation of motion we get:
\begin{equation}
\Omega(r_0)\omega s^2\sim -c_s^2\frac{\delta \Sigma}{\Sigma_0}
\end{equation}
where we find as expected that anticyclonic vortices ($\omega<0$) are associated with high density regions. Combining the velocity profile and the density fluctuation in (\ref{Jdef}) leads to:
\begin{equation}
\label{Jfinal}
\mathcal{J}\sim \Sigma_0 \omega s^3 r_0 \Delta \varphi\Big[-\frac{r_0^2 \Omega(r_0)^2}{c_s^2}+1 \Big]
\end{equation}
where it can be seen that the density fluctuation contribution is larger than the rotation profile contribution by a factor $(r/H)^2\gg 1$. Therefore, a rather surprising result is that anticyclonic vortices are associated with an \emph{excess} of angular momentum within the disk. If a disk region containing a vortex is losing angular momentum by asymmetric wave emission, one deduce from (\ref{Jfinal}) that the vortex can either migrate inward (reduce $r_0$) or reduce its size $s$. The procedure we have used to link wave emission to vortex migration does not allow us to directly compute the migration rate since the vortex size is also expected to change (in particular because $s$ has to be kept smaller than $H$). However, it explains the inward migration observed in simulations of constant vortensity disks. Note that this asymmetry is cancelled out by the effects of a vortensity gradient in the sense that the migration is expected to reverse for a large positive surface density gradient (see section \ref{VORTASM}). In practice migration reversal occurs for $\alpha < \sim 0$ corresponding to a uniform surface density (see  below and Fig. \ref{figmigspeed}). This is less extreme than for the case when the vortex is replaced by  low mass protoplanet which  would migrate inwards and require $\alpha < \sim -1$  when nonlinearity is taken into account \citep{drag} for migration reversal in isothermal disks. Even steeper profiles are needed in the linear migration regime. This because the pressure buffer associated with the fact that an orbiting planet moves at different angular velocity to that of the disk gas when there is a pressure gradient,  which would cause a greater asymmetry in favor of the outer wave,  in fact does not operate for a vortex (see the appendix). Since the vortex is tied to the disk, it will always move with the local gas velocity.
 
\section{Global compressible simulations}
\label{secGlob}

\subsection{Migration for different surface density profiles}

\begin{figure}
\centering
\resizebox{0.8\hsize}{!}{\includegraphics[]{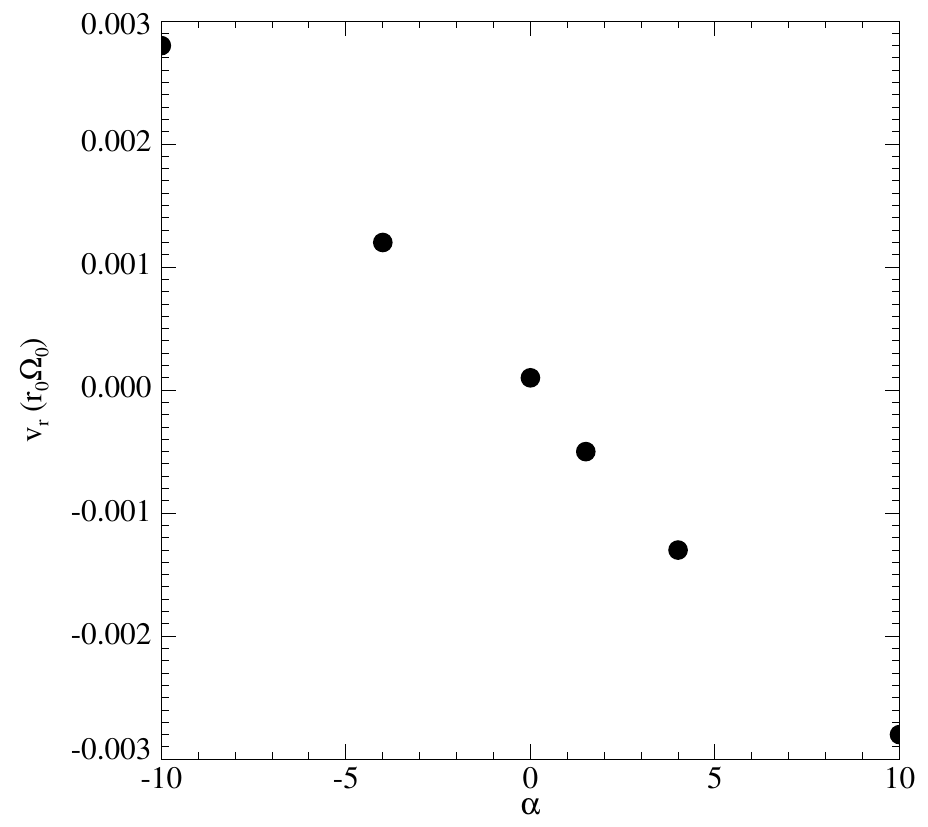}}
\caption{Migration speed in disks with $H_0=0.1r_0$ and the same initial perturbation as in Fig. \ref{figvortdens} for different surface density profiles.}
\label{figmigspeed}
\end{figure}

We have measured the migration speed in isothermal disks with $H_0=0.1r_0$ for various background density gradients; the results are depicted in Fig. \ref{figmigspeed}. The migration speed is approximately proportional to $-\alpha$, with outward migration happening for surface density profiles that increase outward. As indicated above a  constant surface density disk shows almost no vortex migration. For such a  disk, the geometrical effects favoring inward migration (see Sect. \ref{secWave}) are almost exactly balanced by the asymmetry of the vortex, which favors outward migration. The vortex asymmetries make vortex migration more strongly dependent on $\alpha$ than  is the case for linear  Type I planetary migration \citep{tanaka} that requires $\alpha < \sim -2.5$ for migration reversal.
 
\begin{figure}
\centering
\resizebox{0.8\hsize}{!}{\includegraphics[]{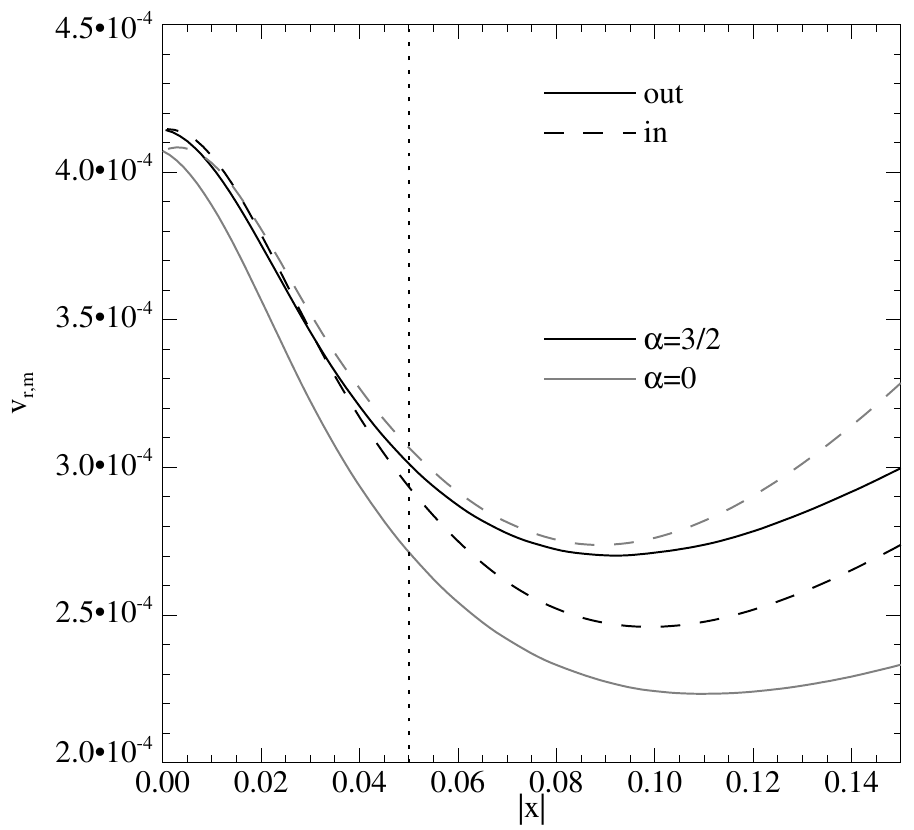}}
\caption{Variation of the $m=10$ component of the radial velocity with distance to the vortex $x=(r-r_\mathrm{vort})/r_\mathrm{vort}$, in an isothermal disk with $H_0=0.1r_0$. The regions inside and outside of the vortex are represented by dashed and solid curves, respectively. Two density profiles are considered: $\alpha=3/2$ (black curves) and $\alpha=0$ (grey curves). The vertical dotted line indicates the approximate extent of the vortex.}
\label{figvortsym}
\end{figure}

The asymmetry in the vortex can be seen for example in the Fourier components of the radial velocity. In Fig. \ref{figvortsym}, we show the $m=10$ component for a constant vortensity disk (black curves) and a constant surface density disk (grey lines). The approximate size of the vortex is indicated by the vertical dotted line. The constant vortensity disk gives rise to a vortex that is almost symmetric for small $|x|$, as expected. The constant surface density disk shows a stronger response in the inner disk (dashed curve), which favors outward migration and works against the geometrical effects of Sect. \ref{secWave}. From Fig. \ref{figmigspeed} we see that for this disk, these competing effects almost cancel each other, resulting in almost no migration for $\alpha=0$. We comment that for a disk with a temperature gradient (through a spatially varying, but constant in time, sound speed), this neutral state corresponds to a disk with constant \emph{pressure}.

\subsection{Migration and disk thickness}

\begin{figure}
\centering
\resizebox{0.8\hsize}{!}{\includegraphics[]{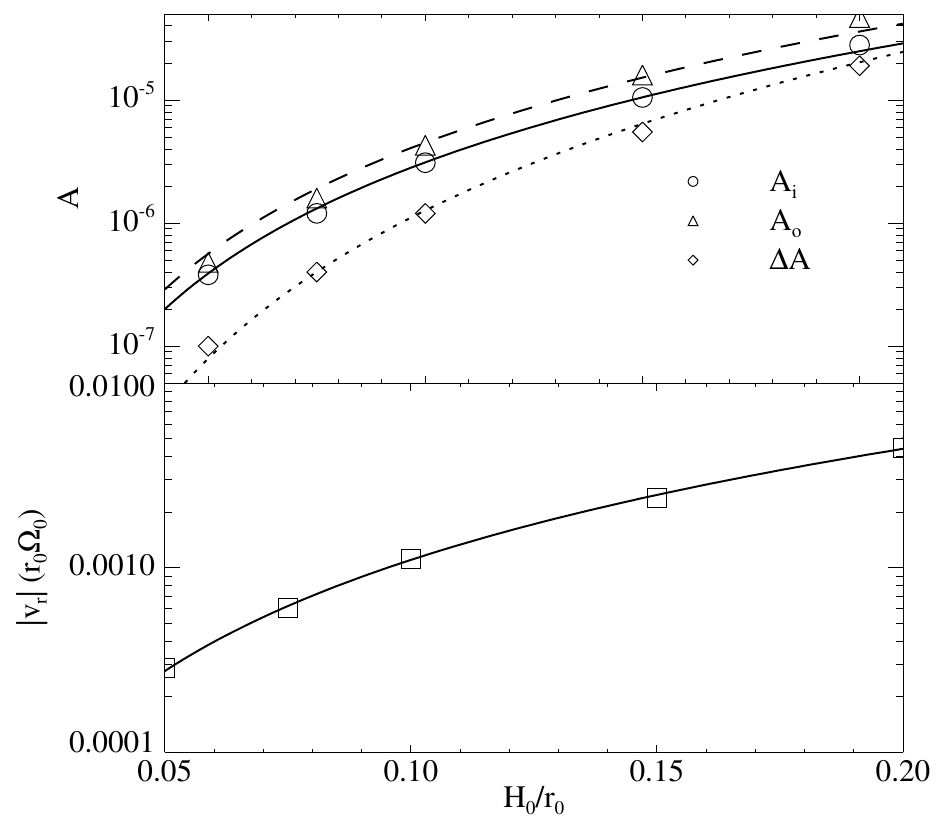}}
\caption{Top panel: wave action due to a vortex of size $H_0/2$, and initial velocity perturbation $3c_s/4$, in disks with $\alpha=3/2$ for different $H_0$. Circles indicate the wave action inside of the orbit of the vortex, triangles the wave action outside the orbit of the vortex, and diamonds the difference between the two. Curves are fits through the data $\propto H_0^3$ (for $A_i$ and $A_o)$, and $\propto H_0^4$ (for $\Delta A$). Bottom panel: corresponding migration speeds. Squares indicate numerical results, the solid curve if a fit through the data $\propto H_0^2$.}
\label{figactionH}
\end{figure}

Until now, we have focused on a disk with $H_0=0.1r_0$. Since we need to be able to resolve the vortex, which is of size $H_0/2$ typically, it is numerically convenient to work with relatively thick disks. However, we expect the angular momentum transport, and therefore the migration speed of the vortex, to depend on $H_0$. From equation (\ref{eqwaveaction}), we expect vortices of size $\sim H_0$ and velocity perturbations $\sim c_s$, that $A \propto H_0^3$. Since the most important contributions to the angular momentum flux come from $m \sim r_0/H_0$, we see from equation (\ref{WASM}) that the ratio of wave actions outside the orbital radius of the vortex and inside the orbital radius of the vortex is $A_o/A_i \approx 1-CH_0/r_0$, with $C$ a numerical constant of order unity. This means that the action difference $\Delta A = A_i-A_o \propto H_0^4$. This equals the rate of angular momentum transfer between vortex and disk, and therefore governs vortex migration.  

In the top panel of Fig. \ref{figactionH} we show the measured wave action inside the orbit of the vortex (circles), outside the orbit of the vortex (triangles), and their difference for different values of $H_0$. The size of the initial perturbation is a fixed fraction ($1/2$) of $H_0$, and the magnitude of the velocity perturbation is a fixed fraction of the sound speed ($3/4$). With these initial conditions, we expect above scalings to hold approximately. We have overplotted fits through the numerical results $\propto H_0^3$ (for $A_i$ and $A_o$) and $\propto H_0^4$ (for $\Delta A$), showing that indeed these scalings hold for a wide range of $H_0$, even in cases where the disk can no longer be called 'thin'. 

The angular momentum of the vortex scales as $H_0^2$, according to equation (\ref{Jfinal}). Combined with an angular momentum transfer rate $\propto H_0^4$ as derived above, we derive a migration time scale $\propto H_0^{-2}$, or a radial velocity of the vortex $\propto H_0^2$. This is the relation we explore in the bottom panel of Fig. \ref{figactionH}, where we show the magnitude of the migration velocity of the vortex for different values of $H_0$. The solid curve is a fit through the numerical data $\propto H_0^2$, showing that again the scaling derived shows good agreement with the numerical simulations. Note that unlike for the case of planet migration, vortex migration is \emph{faster} in thicker disks. This is because the size of the vortex scale with $H_0$, so that for all $H_0$ we get a similar perturbation. This would amount to considering planets of higher mass in thicker disks. 

Even though vortex migration slows down for thinner disks, it is still very substantial at $H_0=0.05 r_0$, a value that is though to be appropriate for protoplanetary disks. From the measured radial velocity, we derive a migration time scale of $r_0/|v_r| = 1700$ orbits. This is very short, but we point out that these results were obtained for a strong vortex with a size comparable to $H_0$. Below, we discuss how migration depends on the size of the vortex.

\subsection{Effects of vortex size and strength}
 
\begin{figure}
\centering
\resizebox{0.8\hsize}{!}{\includegraphics[]{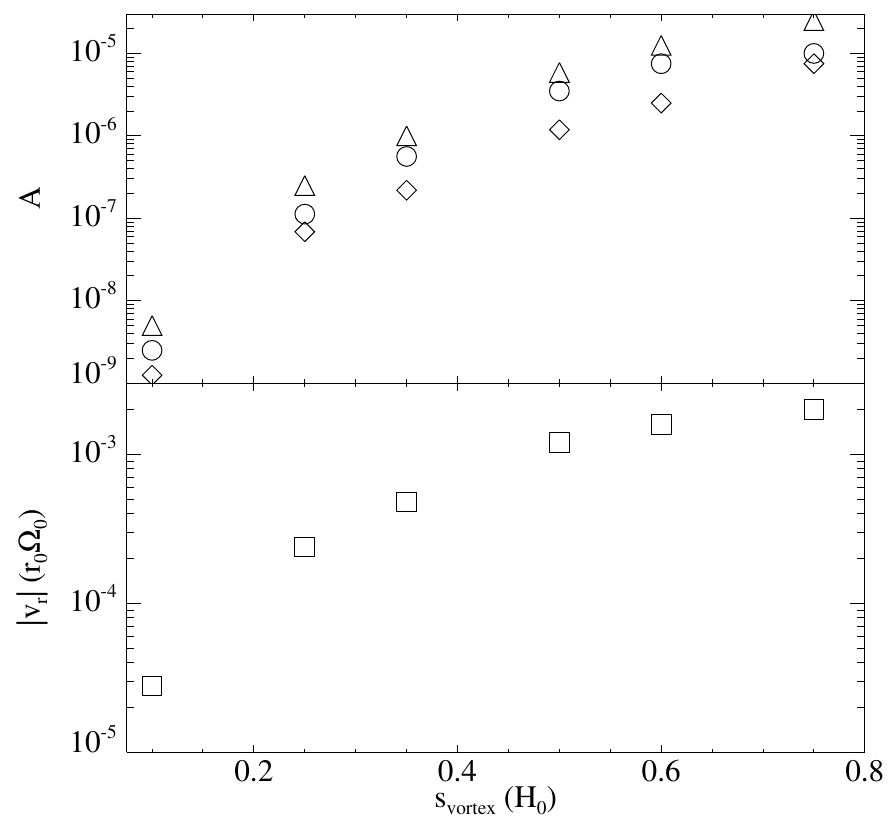}}
\caption{Top panel: wave action due to vortices of different sizes, with initial velocity perturbation $3c_s/4$, in disks with $\alpha=3/2$ and $H_0=0.1 r_0$. Circles indicate the wave action inside of the orbit of the vortex, triangles the wave action outside the orbit of the vortex, and diamonds the difference between the two. Bottom panel: corresponding migration speeds.}
\label{figactionS}
\end{figure}

Vortex migration is driven by perturbations at the sonic line, located roughly at a distance $2H_0/3$ away from the center of the vortex. Away from the core, perturbations due to the vortex decay exponentially in the subsonic region (see equation (\ref{eqattenuation})). We therefore expect smaller vortices to show less migration. Note that a smaller radial vortex size $s$ can be compensated for in principle, as far as migration is concerned, by a stronger initial velocity perturbation.  Unfortunately, the regime $s \ll H_0$ is difficult to study numerically because of computational constraints. We therefore restrict ourselves to $s \geq H_0/10$ to study the effect of vortex size on migration. 

We show numerical results for $H_0=0.1 r_0$ in Fig. \ref{figactionS}. Note that the radial size $s$ quoted in the figure is actually the size of the initial perturbation rather than the actual vortex size. We have always found the two to agree well. We see that indeed smaller vortices migrate slower, with a factor $50$ between a vortex of size $H_0/10$  and a vortex of size $H_0/2$. For $s>H_0/2$, migration is roughly constant, but for the larger cases the vortex extends beyond the sonic line, a situation that is difficult to maintain. 

It is difficult to provide simple scalings in this case, since for most of our results the vortex core occupies a substantial fraction of $H_0$.  We have checked that similar results are obtained for different values of $H_0$. Therefore, even a vortex of size $H_0/10$ with $H_0=0.05 r_0$ would have a migration time scale of $\sim 30000$ orbits only. 

\begin{figure}
\centering
\resizebox{\hsize}{!}{\includegraphics[]{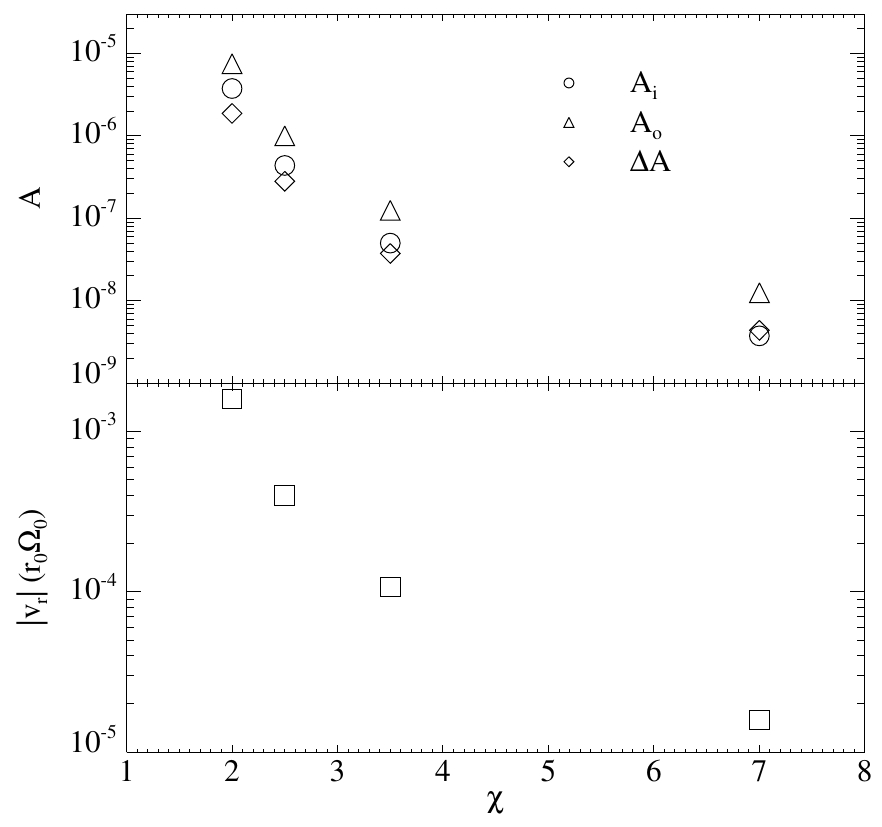}}
\caption{Top panel: wave action due to vortices of different aspect ratio $\chi$, with radial size $H_0/2$, in disks with $\alpha=3/2$ and $H_0=0.1 r_0$. Circles indicate the wave action inside of the orbit of the vortex, triangles the wave action outside the orbit of the vortex, and diamonds the difference between the two. Bottom panel: corresponding migration speeds.}
\label{figactionChi}
\end{figure}

As mentioned above, a stronger velocity perturbation will induce stronger wave emission, and therefore faster migration. For a given radial vortex size, a stronger velocity perturbation will lead to a vortex that is more circular, while weaker vortices have a larger aspect ratio $\chi$. To study the influence of the strength of the vortex, we have varied the initial velocity perturbation at a fixed radial size of the vortex. The aspect ratio of the resulting vortex was measured by considering contours of constant vortensity. 

The resulting wave action and migration speeds are displayed in Fig. \ref{figactionChi}. As expected, stronger vortices migrate faster. This is not only due to a larger perturbation amplitude in the velocity, but also due to a change in the perturbation spectrum. Since the attenuation factor peaks at $m \sim r_0/H_0$, the perturbation amplitudes around this value of $m$ play a major role in determining the migration speed. A vortex of aspect ratio unity and radial size $H_0$ gives rise to large amplitudes at $m \sim r_0/H_0$, and therefore to strong migration. Increasing the aspect ratio shifts the peak of the vortex spectrum to lower values of $m$, which are less effective in inducing angular momentum transfer. This is what is observed in Fig. \ref{figactionChi}. Migration slows down by an order of magnitude when the vortex weakens from $\chi=2$ to $\chi=3.5$. 
    
\subsection{Vortex trapping at surface density maxima}

\begin{figure}
\centering
\resizebox{0.8\hsize}{!}{\includegraphics[]{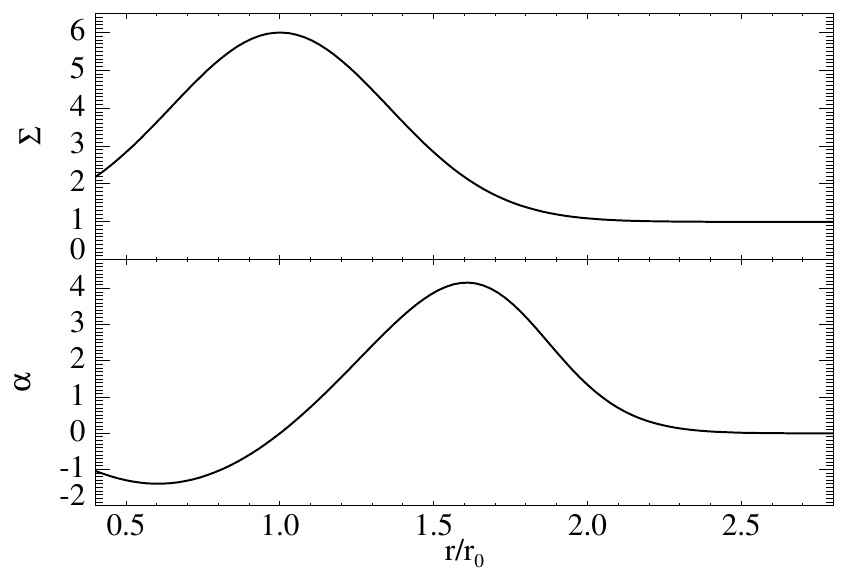}}
\caption{Top panel: surface density profile used to study the stalling of vortices. Bottom panel: the corresponding profile of $\alpha=-d\log \Sigma/d\log r$.}
\label{figbumpdens}
\end{figure}

\begin{figure}
\centering
\resizebox{0.8\hsize}{!}{\includegraphics[]{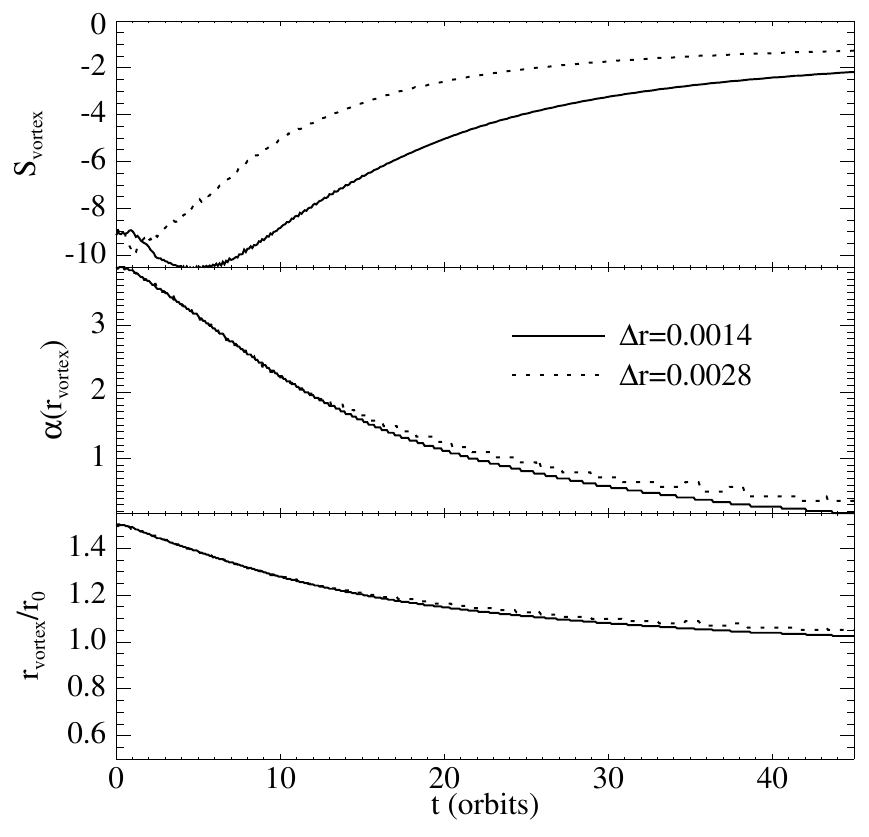}}
\caption{Top panel: evolution of the strength of the vortex, defined by the minimum in the relative vortensity perturbation. Middle panel: the local value of $\alpha$. Bottom panel: orbital evolution of the vortex.}
\label{figbumpmig}
\end{figure}

The strong dependence of migration rates on the background surface density profile opens up the possibility of halting vortex migration at special locations in the disk, where the local gradient of surface density is small. This is explored using a density profile as depicted in the top panel of Fig. \ref{figbumpdens}. The corresponding profile of $\alpha=-d\log \Sigma/ d\log r$ is shown in the bottom panel of Fig. \ref{figbumpdens}. From the discussion above we expect a vortex to migrate inwards outside $r=r_0$, where $\alpha>0$, and outward inside $r=r_0$, where $\alpha<0$. In this configuration, $r=r_0$ is an equilibrium radius for migrating vortices.  

We have put in a strong vortex, with initial velocity perturbation amplitude of $3c_s/4$ over a scale $H_0$, at $r/r_0=1.5$ in a disk with $H_0=0.1r_0$. With such a strong vortex, we make sure it does not dissipate before reaching the equilibrium radius. In Fig. \ref{figbumpmig}, we show the time evolution of the orbital radius of the vortex (bottom panel), the strength of the vortex $S_\mathrm{v}$ (defined by the minimum in the relative vortensity perturbation, top panel), and the local value of $\alpha$ (middle panel). From the top panel we see that the vortex gets weaker with respect to the background flow, partly because of numerical diffusion, partly because of the migration itself. Since the vortex, being a negative perturbation in vortensity,  is moving into a region of lower vortensity, the relative perturbation is expected to go down, as observed in Fig. \ref{figbumpmig}. Results for two different resolutions are shown in Fig. \ref{figbumpmig}; the evolution of $S_\mathrm{v}$ that appears similar in both cases is due to migration, while differences are most likely caused by numerical diffusion. 

From the bottom panel of Fig. \ref{figbumpmig}, we see that migration starts off fast, slows down over time, and almost stalls at $r=r_0$. This slowing down of migration is completely due to the decrease in $\alpha$ as depicted in the middle panel of Fig. \ref{figbumpmig}. At $r/r_0=1.5$, $\alpha\approx 4$, giving rise to fast inward migration. As the vortex approaches the surface density maximum, $\alpha$ approaches zero and migration slows down. Since we obtain essentially the same result for both resolutions, we conclude that the halting of inward migration is due to $r=r_0$ being an equilibrium radius rather than due to the vortex dissipating. This therefore provides an example of vortex trapping at a surface density bump.   
    
\section{Discussion and conclusions}
\label{secDisc}

We have identified a mechanism for angular momentum exchange between a vortex and the surrounding gaseous disk, which leads to orbital migration of the vortex. The vortex excites spiral density waves inside and outside its orbit that will in general be of unequal strength. Geometrical effects favor the outer wave, while a background vortensity gradient in the disk leads to an asymmetric vortex, favoring the wave where the vortensity is highest. For a constant surface density disk, the two effects nearly cancel. 

Whenever the waves are asymmetric, there is an associate angular momentum change in the region around the vortex. The vortex can either shrink, or change its orbital radius. Numerical simulations indicate that vortex migration can be very fast, especially for strong vortices embedded in thick disks, for which we found a migration time scale of a few 100 orbits for $H_0=0.1 r_0$. Migration slows down rapidly for vortices with size $s \ll H_0$. Important differences between vortex migration and Type I planet migration are summarised in Sect. \ref{secCompare} and Table \ref{tabmig}.

A few simplifications were made in the analysis. First of all, we have not solved the energy equation, working with a prescribed fixed sound speed. We do not expect things to change when including the full energy equation. Preliminary simulations show rapid vortex migration in this case as well, with a constant pressure disk again being the neutral state. Migration rates are comparable to the isothermal case.

We have neglected the self-gravity of the disk. Since the vortex is associated with a surface density maximum, it will act as an embedded mass as well as an obstacle. The gravitational interaction of vortex and disk can, completely analogue to the case of an embedded planet, lead to additional migration. However, unless the disk is extremely massive, this effect will be small. This is because a vortex of roughly size $H_0$ perturbs the flow like a planet for which the Hill sphere is comparable in size to $H_0$, or typically a Saturn-mass planet for $H_0=0.05 r_0$. On the other hand, for a Minimum Mass Solar Nebula disk, the mass contained in the vortex is less than an Earth mass. This is one of the reasons why vortex migration can be much more efficient than planet migration: we have an Earth-mass object, emitting density waves like Saturn.

We have worked in the two-dimensional approximation. Although vortices are in general unstable in 3D \citep{lesur09}, they can survive if an excitation mechanism operates \citep[e.g.][]{lesur10}. The two-dimensional approximation is then valid as long as the vertical size of the vortices produced is comparable to $H_0$, since one can then work with vertically averaged quantities. Note however that vortices can also induce large scale 3D circulations (e.g. \citealt{meheut}), in which case the two-dimensional approximation might not be totally valid.

We have considered a single vortex only. Although a detailed description of the interaction of multiple vortices is beyond the scope of this work, we note that in a 2D disk, vortices quickly merge, forming a larger vortex that subsequently migrates faster than the original vortices. 

We have not considered the effects of turbulence in the disk. Our results would therefore apply to a region of the disk where turbulence is almost absent, like a dead zone \citep{gammie}. A thorough discussion of the possible effects of turbulence on vortex migration is beyond the scope of this paper, but we note that if a vortex can survive in a turbulent disk, it would be subject to the same wave torque as discussed in this paper. However, since the wakes are now generated in a stochastic flow, there will be a stochastic component in the torque. For the case of an embedded planet, where the interaction with the turbulence is gravitational, this is known as stochastic migration \citep{nelpap}.

The efficiency of vortex migration raises some important questions about the role of vortices in planet formation. It is well-known that anticyclonic vortices can trap dust particles, and may therefore be efficient sites of planet formation. Now unless vortices are formed at a special location in the disk that is neutral to vortex migration (i.e. a region of constant pressure), the collected solids will rapidly migrate inward with the vortex, all the way to the inner edge of the disk if there is no pressure maximum in between.  Therefore, if vortices play a role in planet formation through accumulating dust particles, all these solids will end up either at a special location in the disk where vortex migration stalls, or at the inner edge of the disk.

In the absence of any stalling location, the life time of a vortex is basically given by its migration time scale, which is only a few thousand orbits for a vortex of size $H_0/2$, aspect ratio $\chi=2$ and $H_0=0.05 r_0$.  This provides a strong time scale constraint for possibly forming long-period (i.e. orbiting at several AU) planets inside a vortex, since they need to decouple from the vortex before reaching the inner edge of the disk. 

Vortices produced by a Rossby wave instability \citep{lovelace} are formed near pressure maxima, and therefore find themselves in a migration neutral location from the start. However, this instability requires strong variations in surface density within one pressure scale height, and it is not clear if such a configuration can be formed naturally in the first place. One possibility is the outer edge of the dead zone \citep{lyra}, but this would require a very sharp transition in resistivity.    

Vortices appear to be naturally produced in global simulations of the MRI \citep{fromang}, and can also grow in regions of a negative entropy gradient by the SBI \citep{lesur10}. In both cases, these vortices would be subject to migration. This way, the outer edge of a dead zone can harbour vortices even in the absence of strong gradients locally, since even a smooth pressure maximum will trap incoming vortices that were generated in the outer disk by either the MRI or the SBI. Note that vortex migration makes it more difficult to sustain the SBI locally. Since the SBI is subcritical, a finite amplitude perturbation is required to start it. Once vortices of size $\sim H_0$ form, they quickly disappear into the inner disk, after which another perturbation is needed to get a second generation of vortices.

In conclusion, vortex migration may be a very effective way of collecting solids in special locations in the disk, such as pressure bumps. Planet formation can then proceed rapidly by coagulation \citep{brauer}, even if the original vortices do not survive. If vortices form readily in protoplanetary disks, this suggests that pressure maxima will be preferred sites for planet formation.
 
\acknowledgments

SJP acknowledges support from STFC in the form of a postdoctoral fellowship. GL acknowledges support from STFC. Simulations were performed using the Darwin Supercomputer of the University of Cambridge High Performance Computing Service (http://www.hpc.cam.ac.uk), provided by Dell Inc. using Strategic Research Infrastructure Funding from the Higher Education Funding Council for England.

 \appendix
 \section{Pressure buffer}
 \label{BUFFER}
We have found the vortex to always orbit at the local gas velocity. This is a different situation to that appropriate to a body that is decoupled from the disk flow, such as a planet,  on which pressure forces do not act, and therefore would move with the local Keplerian angular velocity whereas the gas moves with a \emph{sub-Keplerian} velocity. For the constant vortensity constant aspect ratio disk model discussed in \ref{secWave}, such a body located at $r=r_0$ would rotate with  the Keplerian angular velocity $\lambda\Omega(r_0),$ where $\lambda = 1+5\epsilon^2/4.$ This velocity difference leads to a significantly increased asymmetry in the wave amplitudes that is said to be a consequence of the pressure buffer that slows the rotation of the gas relative to the orbiting object for the disk model considered here. 

It is a simple matter to repeat the calculation in Sect. \ref{secWave} for  the wave attenuation factors appropriate to this case after making the replacement $\omega= m\lambda\Omega(r_0).$ In this case we make the transformation
\begin{equation}
\frac{r}{r_0}=1+\frac{1}{\sqrt{\lambda}}\sqrt{\lambda -1+\epsilon^2+\frac{1}{m^2}}\cos\psi,
\end{equation}
so that the integrals determining them become
\begin{eqnarray}
\ln \Gamma^\pm = -\nonumber \\
\int_{-\pi\frac{-1\pm 1}{4}}^{-\pi\frac{-3\pm 1}{4}}\frac{2m}{3\epsilon}
\frac{(\lambda-1+\epsilon^2+\frac{1}{m^2})\sin^2\psi}{ \sqrt{\lambda}+ \sqrt{\lambda-1+\epsilon^2+\frac{1}{m^2}}\cos\psi}d\psi.
\end{eqnarray}
The  ratio of outward to inward wave flux corresponding to (\ref{WASM})  is then
\begin{equation}
\left(\frac{ \Gamma^+}{\Gamma^-}\right)^2  =\exp \left[- \frac{8 }{9}\sqrt{\frac{\epsilon}{m}}
\left(\frac{9}{4}m\epsilon + \frac{1}{ m\epsilon}\right)^{3/2}\right] .\label{WASMPBF}
\end{equation}
When $m=1/\epsilon$ this leads to a ratio $1-5.2/m,$ corresponding to an asymmetry about twice as large as given by (\ref{WASM}). Therefore, the pressure buffer is an important additional source of asymmetry in the case of an embedded object on a Keplerian orbit such as a planet \citep[see also][]{ward}.
\bibliography{vortex}

\end{document}